\documentclass[utf8]{frontiersSCNS} 

\usepackage{microtype,subcaption}
\usepackage[onehalfspacing]{setspace}
\usepackage{graphicx}
\usepackage{epstopdf}
\usepackage{mathrsfs}
\usepackage{amsmath}
\usepackage{natbib}
\usepackage{color}
\usepackage{changes}
\usepackage{hyperref}

\renewcommand{\emph}{\textit}

\newcommand{\pD}[2]{\frac{\partial #2}{\partial #1}}
\newcommand\bb[1]{\mbox{\boldmath{$#1$}}}
\newcommand\grad{\bb{\nabla}}
\newcommand\bcdot{\,\bb{\cdot}\,}
\newcommand\btimes{\,\bb{\times}\,}
\newcommand\ldquo{`}
\newcommand\rdquo{'}
\newcommand\sci{Science}
\newcommand\nat{Nature}
\newcommand\natph{Nat.~Phys.}
\newcommand\prl{Phys.~Rev.~Lett.}
\newcommand\grl{Geophys.~Res.~Lett.}
\newcommand\jgr{J.~Geophys.~Res.}
\newcommand\jgra{J.~Geophys.~Res.~A}
\newcommand\njp{New~J.~Phys.}
\newcommand\jpp{J.~Plasma~Phys.}
\newcommand\pop{Phys.~Plasmas}

\newcommand\pofb{Phys.~Fluids~B}
\newcommand\apj{Astrophys.~J.}
\newcommand\apjl{Astrophys.~J.~Lett.}

\newcommand\jcomp{J.~Comput.~Phys.}

\def\keyFont{\fontsize{8}{11}\helveticabold }
\def\firstAuthorLast{Califano et al.} 
\def\Authors{F.~Califano$^{1}$, S.~S.~Cerri$^{2}$, M.~Faganello$^{3}$, D.~Laveder$^{4}$, M.~Sisti$^{1,3}$, M.~W.~Kunz$^{2,5}$}
\def\Address{$^1$Dipartimento di Fisica `E. Fermi', Universit\'a di Pisa, Pisa, Italy\\
$^2$Department of Astrophysical Sciences, Princeton University, Princeton, NJ 08544, USA\\
$^3$Aix-Marseille Universit\'e, CNRS, PIIM UMR 7345, 13397, Marseille, France\\
$^4$Universit\'e C\^ote d'Azur, CNRS, Observatoire de la C\^ote d'Azur, Laboratoire J. L. Lagrange, Boulevard de l'Observatoire, CS 34229, 06304 Nice  Cedex 4, France\\
$^5$Princeton Plasma Physics Laboratory, PO Box 451, Princeton, NJ 08543, USA}
\def\corrAuthor{}
\def\corrEmail{francesco.califano@unipi.it; scerri@astro.princeton.edu; matteo.faganello@univ-amu.fr}

\begin{document}
\onecolumn
\firstpage{1}

\title[e-rec in plasma turbulence]{Electron-only reconnection in plasma turbulence} 

\author[\firstAuthorLast ]{\Authors} 
\address{} 
\correspondance{} 
\extraAuth{}

\maketitle

\begin{abstract}
Hybrid-Vlasov--Maxwell simulations of magnetized plasma turbulence including non-linear electron-inertia effects in a generalized Ohm's law are presented. When fluctuation energy is injected on scales sufficiently close to
ion-kinetic scales, the ions efficiently become de-magnetized and electron-scale current sheets largely dominate the distribution of the emerging current structures, in contrast to the usual picture, where a full hierarchy of structure sizes is generally observed. These current sheets are shown to be the sites of electron-only reconnection (e-rec), in which the usual electron exhausts are unaccompanied by ion outflows and which are in qualitative agreement with those recently observed by {\it MMS} in the Earth's turbulent magnetosheath, downstream of the bow shock. 
Some features of the e-rec phenomenology are shown to be consistent with an electron magnetohydrodynamic description. Simulations suggest that this regime of collisionless reconnection may be found in turbulent systems where plasma processes, such as micro-instabilities and/or shocks, overpower the more customary turbulent cascade by directly injecting energy close to the ion-kinetic scales.

\tiny
 \keyFont{ \section{Keywords:} magnetic fields, magnetic reconnection, plasma turbulence, solar wind, Earth magnetosheath, plasma simulations}
 
\end{abstract}

\section{Introduction}\label{sec:intro}

Highly accurate {\it in situ} satellite measurements of plasma fluctuations and particle distribution functions in the heliosphere are today the primary experimental tool for the study of plasma turbulence~\citep[e.g.][]{AlexandrovaPRL2009,SahraouiPRL2009,ChenBoldyrev2017} and magnetic reconnection~\citep[e.g.][]{RetinoNAT2007,BurchSCI2016}. In particular, space missions such as {\it Cluster} and {\it MMS} provide unprecedented opportunities to investigate multi-scale plasma physics, from the ion-kinetic scales down to the electron-kinetic scales, and thereby to constrain theoretical models of kinetic turbulence and reconnection.

Recently, {\it MMS} has measured in the Earth's turbulent magnetosheath a series of electron-scale reconnection events in which super-Alfv\'{e}nic electron jets are never accompanied by ion outflows \citep{PhanNAT2018}. 
Those authors dubbed these events `electron-only reconnection' (hereafter, e-rec), to differentiate them from the usual (fast) collisionless reconnection process  \citep[e.g.,][]{coppi64,drake91,Ottaviani93,DrakeGRL1997} in which reconnection takes place in an `electron diffusion region' embedded within a larger `ion diffusion region' and in which both species are expelled in a collimated outflow named the `exhaust' \citep[e.g.,][]{ShayJGR1998}. While e-rec has been recently shown to occur in kinetic simulations of magnetic reconnection when a single electron-scale CS is ad-hoc initialized \citep{SharmaPyakurel2019} or after a quasi-parallel shock \citep{Bessho2019}, it is difficult to explain e-rec in the context of a classical turbulent cascade, in which turbulent energy is transferred conservatively from magneto-fluid scales down to ion-kinetic scales, thereby coupling to the ion dynamics in the usual way~\citep{RetinoNAT2007,SundkvistPRL2007,ServidioPRL2012,PerroneAPJ2013,PerronePOP2018,KarimabadiPOP2014,FranciAPJ2016,FranciAPJL2017,WanPOP2016,CerriCalifanoNJP2017,CerriJPP2017,ShayPOP2018}. 
This difficulty also holds in scenarios in which the turbulent cascade of magnetic fluctuations is shown to be mediated by the reconnection of ion-scale current sheets~\citep[e.g.,][]{CerriCalifanoNJP2017,FranciAPJL2017,LoureiroBoldyrevAPJ2017,MalletJPP2017,PapiniAPJ2019}. 
Indeed, to the best of our knowledge, simulations in which energy is injected at scales much larger than the ion-kinetic length scales ($d_{\rm i}$, $\rho_{\rm i}$ being of the same order) show reconnecting CSs with {both electron and ion outflows}, regardless of the nature of this injection -- whether it be continuous forcing or an initial distribution of magnetic and/or velocity fluctuations. This is true also for satellite observations of heliospheric turbulence (now able to resolve the electron scales; e.g.~\citet{VorosJGR2017}), except for those {\it MMS} observations cited above. 

In this paper, we use hybrid-Vlasov--Maxwell (HVM) simulations of freely decaying, 2D-3V turbulence to show that e-rec events qualitatively similar to those observed by {\it MMS} can develop in an `MHD-scale' turbulent system if magnetic fluctuations are injected sufficiently close to the ion-kinetic scale.
These simulations include the Hall and electron-inertia terms in a generalized Ohm's law~\citep{ValentiniJCP2007} that 
captures the decoupling of the magnetic field from the ion dynamics at ion-kinetic scales and allows the complete unfreezing of magnetic flux with respect to the electron fluid motion at the electron-inertial scale. We also demonstrate that simulations of plasmas under the same conditions but with turbulent energy injected farther away from the ion-kinetic scale do not show e-rec, and instead exhibit  standard multi-scale reconnection with both ion and electron outflows. The transition from standard reconnection to e-rec is found to occur when the wavenumber range of the injected fluctuations gets close to the ion-kinetic scales, namely when the largest injected wavenumber is varied from $(k_\perp d_\mathrm{i})_\mathrm{max} = 0.3$ to $(k_\perp d_\mathrm{i})_\mathrm{max} = 0.6$. 
Finally, we show that the physics underlying some features of the e-rec phenomenology can be described by the equations of electron magnetohydrodynamics~\citep[EMHD;][]{KingsepVTP1987,BulanovPOF1992,MandtGRL1994}.

\section{Method of Solution}

\subsection{Basic equations}
We integrate the Vlasov equation for the ion distribution function $f_{\rm i}(t,\bb{r},\bb{v})$ using an Eulerian approach~\citep{MangeneyJCP2002}, coupled with Faraday's law of induction for the evolution of the magnetic field $\bb{B}(t,\bb{r})$. In dimensionless units, these are, respectively,
\begin{equation}\label{eq:Vlasov}
\pD{t}{f_{\rm i}}\,
+\bb{v}\bcdot\grad f_{\rm i}\,
+\bigl(\bb{E}+\bb{v}\btimes\bb{B}\bigr)\bcdot\pD{\bb{v}}{f_{\rm i}} = 0 ,
\end{equation}
\begin{equation}\label{eq:Faraday}
\pD{t}{\bb{B}} = -\grad\btimes\bb{E} .
\end{equation}
Within the hybrid-Vlasov--Maxwell (HVM) framework, these equations are coupled to an Ohm's law for the electric field $\bb{E}(t,\bb{r})$ that includes contributions from the pressure and inertia of an isothermal electron fluid~\citep{ValentiniJCP2007}:
\begin{equation}\label{eq:TF_ohm}
\bigl(1-d_{\rm e}^2\nabla^2\bigr)\bb{E} = -\bb{u}_{\rm e}\btimes\bb{B} - T_{\rm e}\grad\ln n + d_{\rm e}^2 \,\grad\bcdot\bigl[n(\bb{u}_{\rm i}\bb{u}_{\rm i}-\bb{u}_{\rm e}\bb{u}_{\rm e})\bigr].
\end{equation}
%
All equations are normalized with respect to the ion quantities: the mass $m_{\rm i}$, charge $e$, inertial length $d_{\rm i}$, and cyclotron frequency $\Omega_{\rm i} \doteq eB_0/m_i c$, where $B_0$ is the strength of the mean magnetic field. The electron fluid is characterized by a finite electron skin depth $d_{\rm e}=\sqrt{m_{\rm e}/m_{\rm i}}$, a flow velocity $\bb{u}_{\rm e}=\bb{u}_{\rm i}-\bb{J}/n$, and a constant electron temperature $T_{\rm e}$. The number density $n$ and the ion momentum density $n\bb{u}_{\rm i}$ are computed as the zeroth and first velocity-space moments of $f_{\rm i}$, respectively. Quasi-neutrality is assumed, $n_{\rm e}\simeq n_{\rm i} = n$, and the displacement current is neglected in Amp\'ere's law, so that the current density $\bb{J} = \grad \btimes \bb{B}$.
Electron inertia, see Eq.~(\ref{eq:TF_ohm}), is a key ingredient in this work as it allows for collisionless reconnection to occur even in the framework of a fluid-electron model. At the same time, the Hall term (hidden in the $-{\bf u}_e \times {\bf B}$ term, equivalent to $-{\bf u}_i \times {\bf B} + {\bf  J}/n \times {\bf B}$) is the key actor for speeding up the dynamics around the X-point and leading to ``fast'' magnetic reconnection (see \cite{Birn2001} or Appendix B in \cite{faganello09} for further details). 
We mention that a fully kinetic approach would be more appropriate for the physical description of the diffusive effects since the agyrotropic electron pressure tensor plays a major role (see for instance \cite{cai97, hesse99}. However, such an approach is out of the scope of this paper and left for future investigation. \\

\subsection{Simulation setup}
\label{sec:sims}

We consider an initially uniform, Maxwellian, proton--electron plasma embedded in a homogeneous out-of-plane magnetic field, $\bb{B}_0 = B_0\hat{\bb{e}}_z$. The plasma beta parameter for the ion species initially satisfies $\beta_{\rm i}\doteq 8\pi n T_{\rm i0}/B^2_0 = 1$, where $T_{\rm i0}$ is the initial ion temperature. We set $T_{\rm e} = T_{\rm i0}$. A reduced mass ratio of $m_{\rm i}/m_{\rm e}=144$ is employed, ensuring that $d_{\rm i}$ and $d_{\rm e}$ are well separated.
The HVM equations (\ref{eq:Vlasov})--(\ref{eq:TF_ohm}) are then solved in a 2D-3V phase space with $1024^2$ grid points spanning a real-space domain of size $L_x=L_y = 20\pi d_{\rm i}$, corresponding to a uniform spatial resolution ${\simeq}0.06 d_{\rm i} \simeq 0.7d_{\rm e}$. 
In order to avoid spurious numerical effects at very small scales, we adopt high-order spectral filters \citep{LeleJCP1992} on the electromagnetic fields that act only on the high-k part of the spectrum for numerical stability, while reconnection is driven by electron inertia.
The velocity space is sampled by $51^3$ uniformly distributed grid points spanning $[-5,5]v_{\rm th,i}$, where $v_{\rm th,i}=\sqrt{\beta_{\rm i}/2}$ is the initial ion thermal speed, corresponding to a resolution of $0.2 v_{\rm th,i}$ in each velocity direction.
We note that velocity fluctuations smaller than the grid resolution, $\Delta v = 0.2$ (in normalized units) in all directions, are well recovered by the Eulerian approach thanks to the very low level of noise of the algorithm. Indeed, even when the mean flow is small (with respect to $\Delta v$) the asymmetries in the values that the distribution function assumes on the negative and positive part of the v-space grid are well captured by its first-order moment $\int \bb{v} f(\bb{v}) \mathrm{d}^3\bb{v}$. We caution that this may not be true only if the width of the distribution is not well resolved, i.e., if $\Delta v/v_\mathrm{th} > 1$.

We note that velocity fluctuations slower than the grid resolution, dv = 0.2 in all directions, are well recovered by the Eulerian approach where the full sampling of the distribution function allows to capture any mean-flow asymmetry except for the cold plasma limit.

Decaying turbulence is initialized by imposing random, isotropic magnetic-field perturbations, $\delta\bb{B} = \delta B_\parallel \hat{\bb{e}}_z + \delta\bb{B}_\perp$ with $\grad\bcdot\delta\bb{B} = 0$. Such perturbations self-consistently excite `compressive' fluctuations in the velocity and density fields~\citep{CerriJPP2017}. The choice of injecting compressive magnetic perturbations (rather than, for instance, purely Alfv\'enic fluctuations having $\delta B_\parallel = 0$) is justified by the fact that inside the bow shock (i.e.~in the magnetosheath) the majority of the fluctuations are observed to be compressive~\citep[see, e.g.,][]{HuangAPJL2017,HadidPRL2018}. Furthermore, the injection scale in our simulations resides (purposefully) close to $d_{\rm i}$, at which `Alfv\'enic' or `magneto-sonic' fluctuations are not clearly distinguishable. Accordingly, we let the (kinetic) response of the plasma to the magnetic-field perturbations to develop velocity and density fluctuations self-consistently. 
This particular choice has been indeed shown to not affect the development of the turbulent cascade at (and below) the ion scales~\citep{CerriJPP2017}. However, a study about the precise dependence on a much wider variety of injection properties is beyond the scope of this work -- which itself is meant to be a first proof-of-concept -- and will be left for future investigation.

Two simulations, labeled sim.A and sim.B for convenience, have been performed. In sim.A, the wavenumbers $k$ of the injected magnetic-field perturbations lie in the range $0.1\leq (k_\perp d_{\rm i})_\mathrm{inj} \leq 0.6$ with a corresponding rms amplitude of the in-plane perturbations  $|\bb{e}_z\btimes\grad\psi|^{\rm rms}\simeq0.15 $. These values are chosen so that moderate-amplitude fluctuations are injected on scales close to the ion-kinetic scales, so as to separate the ion and the electron dynamics from the very beginning of the simulation (see the discussion in \S\,\ref{subsec:EMHDinHVM}). 
In sim.B we excite fewer modes whose wavenumbers lie farther away from the ion-kinetic scales, {\it viz.}, $0.1\leq (k_\perp d_{\rm i})_\mathrm{inj} \leq 0.3$, corresponding to the regime studied in \citet{FranciAPJ2015, FranciAPJL2017}, \citet{CerriAPJL2016,CerriJPP2017}, and \citet{CerriCalifanoNJP2017}. 
We adopt the same rms value of the magnetic fluctuations as in sim.A.
A third simulation, sim.C, which is equivalent to sim.B in terms of injection scales ({\it viz.}, $0.1\leq (k_\perp d_{\rm i})_\mathrm{inj}\leq0.3$) but employs a larger rms value for the fluctuations ({\it viz.}, $|\bb{e}_z\btimes\grad\psi|^{\rm rms}\simeq0.21$), has also been performed. It confirms that the qualitative results from sim.B do not depend strongly on the initial level of the fluctuations, although a more precise statement will require further studies that are left for future investigation. 
Therefore, since there is no more information in sim.C that is not already present in sim.B, in this paper we only present results from sim.A and sim.B. \\

\subsection{The EMHD limit at sub-ion scales}\label{subsec:EMHDinHVM}

As turbulent fluctuations cascade to smaller and smaller scales, their characteristic frequencies increase, eventually exceeding those described by the MHD model. At scales below the ion-kinetic scales but above the electron skin depth, the ions decouple from the magnetic field, which remains frozen in the electron fluid. In the limit where the spatial derivatives of the number density $n$ and of the ion bulk flow $\bb{u}_{\rm i}$ can be neglected with respect to the derivatives of the electron fluid velocity $\bb{u}_{\rm e}$, the plasma may be described by EMHD. In this case, equations (\ref{eq:Faraday}) and (\ref{eq:TF_ohm}) reduce to equation (7) of \citet{BulanovPOF1992}, in which the field $\bb{B}'\doteq\bb{B}-d_{\rm e}^2\nabla^2\bb{B}$ (rather than $\bb{B}$) is frozen into an incompressible electron flow and the current is carried solely by the electron fluid, so that $\bb{u}_{\rm e} = -\bb{J}/n$. Under these conditions,  collisionless EMHD in two dimensions ensures the Lagrangian conservation of $F\doteq\psi-d_{\rm e}^2\nabla^2\psi$ \citep{morrison80,BulanovPOF1992}, where the flux function $\psi$ is related to the magnetic field via  $\bb{B}=\bb{e}_{z}\btimes\grad\psi+\bb{B}_z$. Note that while $\bb{B}'$ and $F$ stay frozen in the electron fluid, $\bb{B}$ and $\psi$ can reconnect. One advantage of the HVM model adopted here is that it includes EMHD as one of the limits of the generalized Ohm's law (\ref{eq:TF_ohm}), while simultaneously accounting also for non-EMHD effects (e.g., compressibility) and the kinetic response of the ions.
On this point, we also note that the electron inertia terms in Eq. \ref{eq:TF_ohm}, aimed first at allowing reconnection without resistivity, should include a term proportional to $d_e^2 \nabla(\nabla \cdot {\bf E})$. Even if it is negligible in the EMHD limit, this term could influence the kinetic ion response since the electric field directly enters the ions' Vlasov equation.
Our set of equations can be seen as the simplest model describing kinetic ions, fluid electrons, ion and electron decoupling at different scales without including dissipation, finally conserving the EMHD flux $F$ in the high-frequency limit. 
We nevertheless underline that in Eq.~(\ref{eq:TF_ohm}) for the electric field we do not formally impose $\nabla \cdot E = 0$. As a consequence the electrostatic feedback is anyway included in our hybrid model (see for instance the test on  Landau damping on ion acoustic waves in \cite{ValentiniJCP2007}).
\begin{figure*}
\centering
\includegraphics[width=0.5\columnwidth]{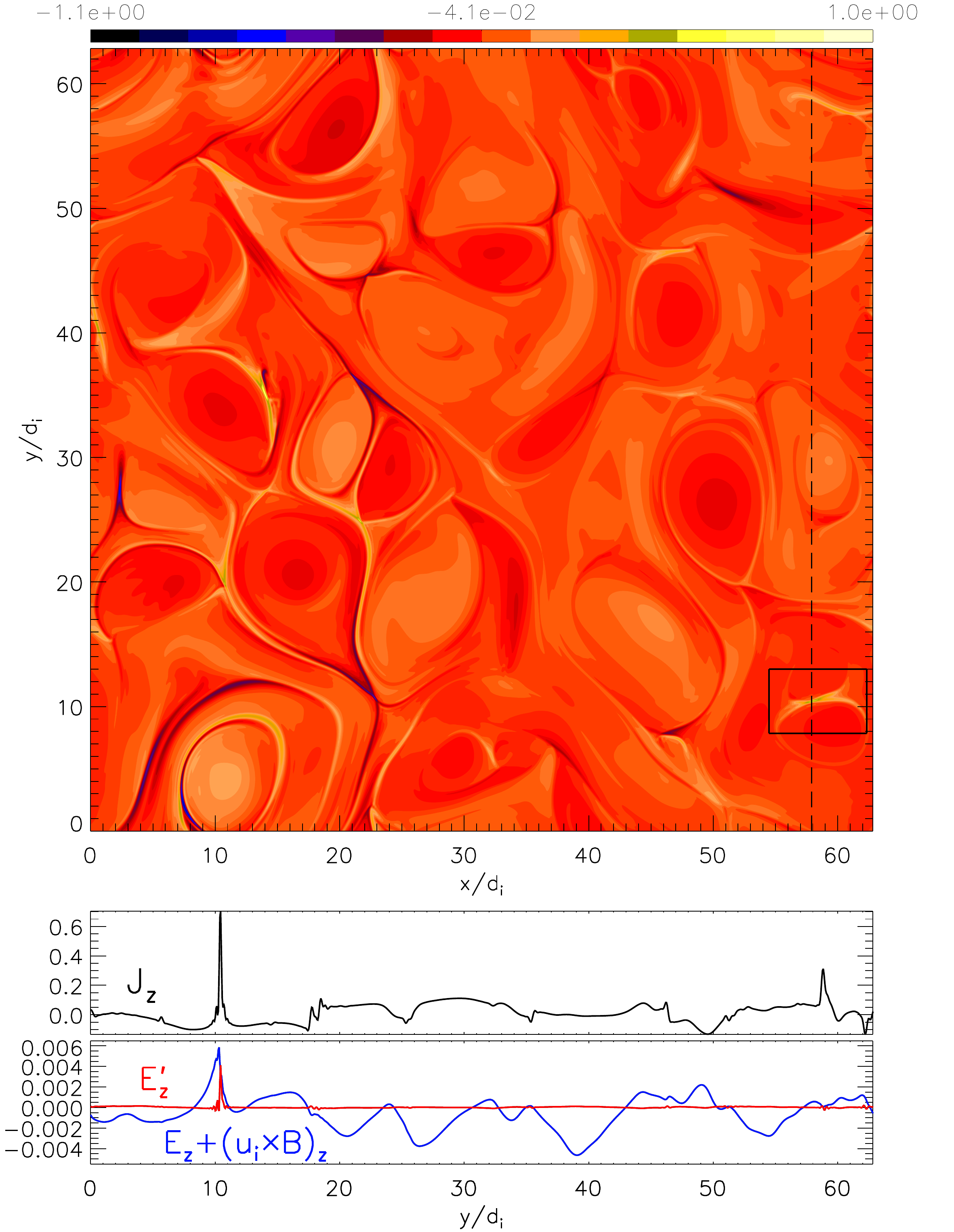}%
\includegraphics[width=0.51\columnwidth]{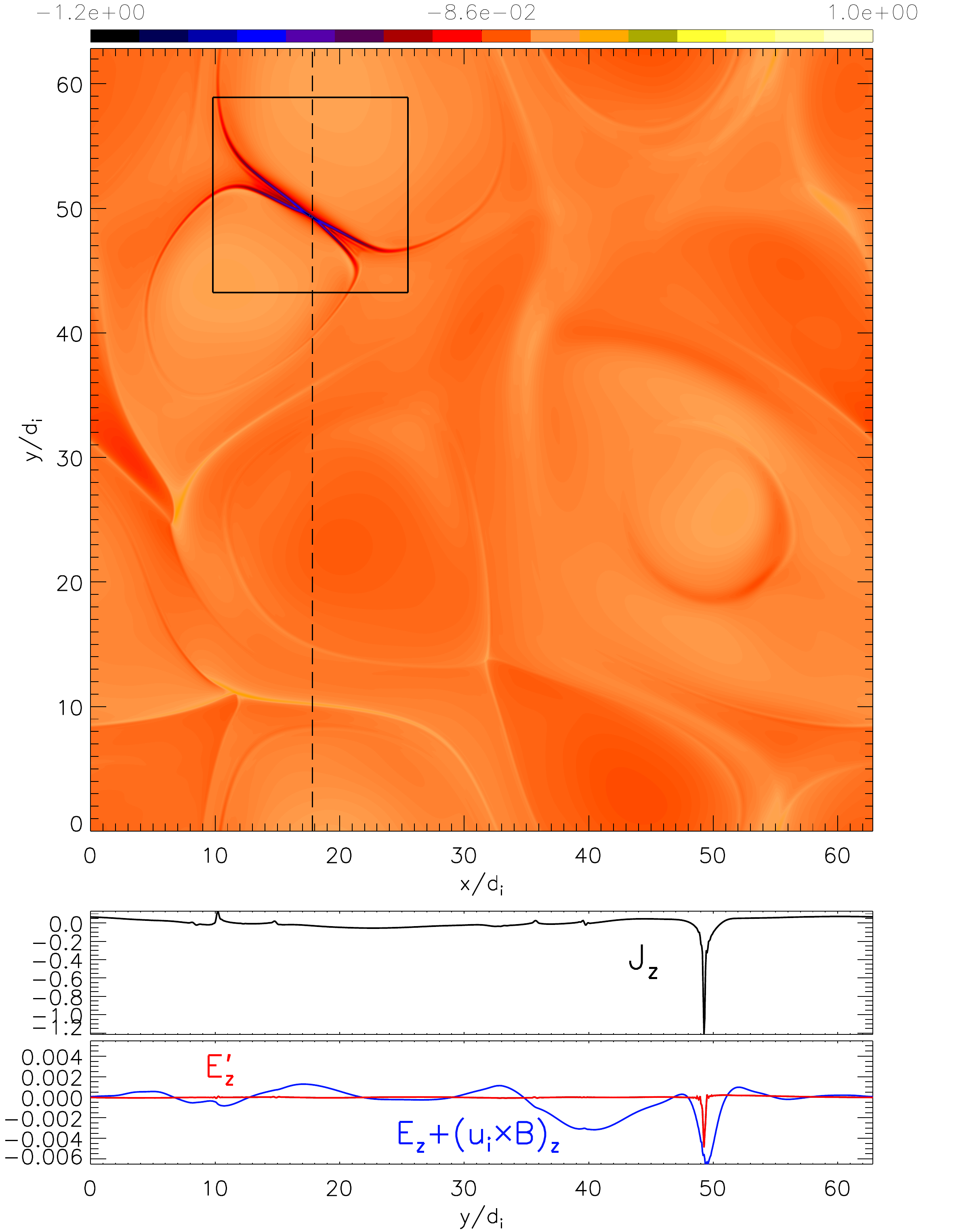}
\caption{Left: From sim.A. (top) Shaded iso-contours of the out-of-plane current density $J_z$ at $t=42.5\Omega_{\rm i}^{-1}$. The rectangle indicates the position of CS1. (bottom) One-dimensional cut along $y$ at $x\simeq58$ (marked by the vertical dashed line in the top panel) of $J_z$ (black line), $E_z'\doteq{E_z}+(\bb{u}_{\rm e}\btimes\bb{B})_z$ (red line), and $E_z+(\bb{u}_{\rm i}\btimes\bb{B})_z$ (blue line). 
Right: From sim.B. (top) Shaded iso-contours of the out-of-plane current density $J_z$ at $t=95\Omega_{\rm i}^{-1}$. 
The square indicates the CS shown in figure \ref{fig6} (the simulation plane has been slightly shifted to highlight the two magnetic islands forming this CS).
(bottom) One-dimensional cut along $y$ at $x\simeq18$ (marked by the vertical dashed line in the top panel) of $J_z$ (black line), $E_z'\doteq{E_z}+(\bb{u}_{\rm e}\btimes\bb{B})_z$ (red line), and $E_z+(\bb{u}_{\rm i}\btimes\bb{B})_z$ (blue line).
(Note that the $y$-range of the bottom plot for $E_z'$ has just been shifted with respect to the corresponding plot for sim.A because the current peak in sim.B is negative. However, the length of the $y$-range shown, $|y_{\rm max}-y_{\rm min}|$, is the same to facilitate a meaningful comparison.)}
\label{fig1} 
\end{figure*}
Numerically, we have two points to solve the electron inertial length $d_e$ which is a somewhat borderline to completely separate the EMHD invariant from the flux function. This means that numerical filtering or intrinsic dissipation of the algorithm to advance the Vlasov equation (see \cite{MangeneyJCP2002}), could contribute to the reconnection process and partially dissipate $F$ and so $\psi$. We note (see Section \ref{sec:simA}, in particular Fig.\ref{fig2}) that $F$ and $\psi$ are quite well separated during their evolution suggesting that, even at a resolution that must menage at the same time the large-scale evolution and the small-scale reconnection events, electron inertia terms play an important role in allowing for reconnection to occur.

\section{Results}\label{results}

\subsection{Sim.A: Signatures of e-rec in kinetic turbulence}\label{sec:simA}

We first describe the results from sim.A. 
Within roughly $1.5$ eddy turnover times of the shortest initial wavelength (${\sim}30\Omega_{\rm i}^{-1}$), CSs form and reconnection onsets. (Fully developed turbulence is realized only on the eddy turnover time of the outer-scale fluctuations, $\tau\sim100\Omega_{\rm i}^{-1}$.) In figure \ref{fig1}, top left, we show a full-box view of the out-of-plane current density $J_z$ at $t=42.5\Omega_{\rm i}^{-1}$. For comparison, in the top right frame, we show the same quantity from sim.B (discussed in \S\,\ref{sec:simBC}). We observe the formation of several electron-scale CSs, e.g., CS1 (indicated by the rectangle box) at $(x,y)=(58, 11)$, CS2 at $(23, 11)$, and CS3 at $(21, 36)$. These CSs are characterized by widths {of a few $d_{\rm e}$}.
More interestingly, their lengths $L_{\rm CS}\sim (2$--$5) d_{\rm i}$ are shorter than those typically observed in other turbulence simulations \citep[e.g.,][]{CerriJPP2017,CerriCalifanoNJP2017,FranciAPJ2016,FranciAPJL2017,ServidioJPP2015,ValentiniPOP2014}.
On these scales, the electron fluid, and thus the magnetic field, is expected to decouple from the ions; it is only deep within these thin CSs that the magnetic field ultimately decouples from the electron fluid. These features are shown in figure \ref{fig1}, bottom frame, which displays $J_z$ (black), the out-of-plane electric field in the frame of the electrons $E_z'\doteq{E_z}+(\bb{u}_{\rm e}\btimes\bb{B})_z$ (red), and the out-of-plane electric field in the frame of the ions $E_z+(\bb{u}_{\rm i}\btimes\bb{B})_z$ (blue), both versus $y$ at $x\approx 58d_{\rm i}$ (corresponding to the vertical dashed line in the accompanying top panel). While the electron fluid is seen to decouple from the magnetic field only deep within CS1 (at $y\approx 11d_{\rm i}$, where $E'_z\not\approx{0}$), the ions are approximately decoupled over much of the simulation domain (i.e., $E_z+(\bb{u}_{\rm i}\btimes\bb{B})_z\not\approx{0}$ at all scales). The width of CS1 (either inferred by the condition $|J_z|>J_z^{\rm rms}$ or by using the more accurate method discussed in \S\,\ref{sec:manu}) is $\ell\approx 4d_{\rm e}$, corresponding to the physical range on which e-rec has been observed to occur in the magnetosheath~\citep{PhanNAT2018}.

\begin{figure}
\centering
\includegraphics[width=0.65\columnwidth]{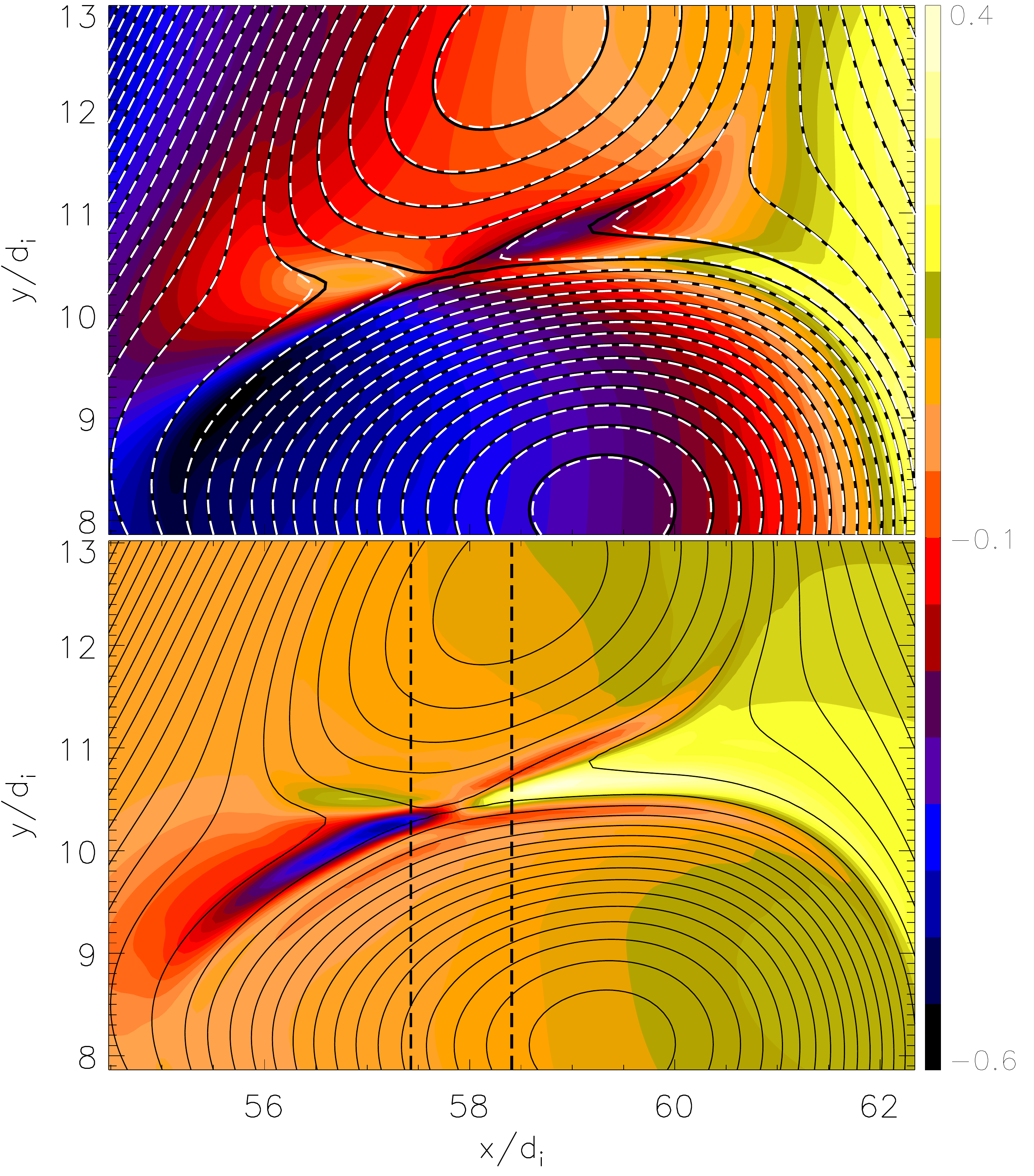}
\caption{From sim.A. Zoom-in on CS1 at $t=42.5\Omega_{\rm i}^{-1}$. (top) Shaded iso-contours of the out-of-plane magnetic field $B_z$ with superposed iso-contours of $\psi$ (dashed white) and $F$ (black). (bottom) Shaded iso-contours of the $x$-component of the electron flow, $u_{{\rm e},x}$, and of $\psi$ (solid black lines). Dashed vertical lines trace the two virtual spacecraft trajectories shown in the two columns of figure \ref{fig3}.}\label{fig2}
\end{figure}

Figure \ref{fig2} presents a zoom-in of CS1 after the onset of e-rec and the formation of an active X-point structure at $t=42.5\Omega^{-1}_{\rm i}$. The top panel reveals a quadrupolar structure of $B_z$ near the CS, known to be caused by the Hall term allowing the electrons to locally decouple from the ion neutralizing background and to generate via their in-plane current the quadrupole ~\citep{MandtGRL1994,Biskamp97,ShayJGR1998,UzdenskyKulsrudPOP2006}.
Quadrupole signature is today a kind of universal trademark of fast reconnection routinely used not only in simulations but also in space observations to assert the presence of a reconnection event~ \citep{Oieroset2001,vaivadsPRL2004}.
Iso-contours of the flux function $\psi$ (dashed white lines) and of the corresponding EMHD invariant $F$ (black lines) coincide on scales larger than $d_{\rm e}$, where electron inertia is unimportant. Both quantities are advected by the flow and well conserved everywhere except around the X-point, where $\psi$ locally breaks (and reconnects) on scales ${\sim}d_{\rm e}$. 
This separation of the $\psi$ and $F$ iso-contours is a signature of the EMHD regime~\citep[see, e.g.,][]{AtticoPOP2002}. Indeed, in the bottom panel of figure \ref{fig2}, the $x$-component of the electron flow $u_{{\rm e},x}$ shows the typical electron jet structures coming out from the X-point (highlighted by $F$, the solid black lines). No evidence of corresponding ion outflows is found in the exhaust around the X-point.

\begin{figure*}
\centering
\includegraphics[width=0.8\columnwidth]{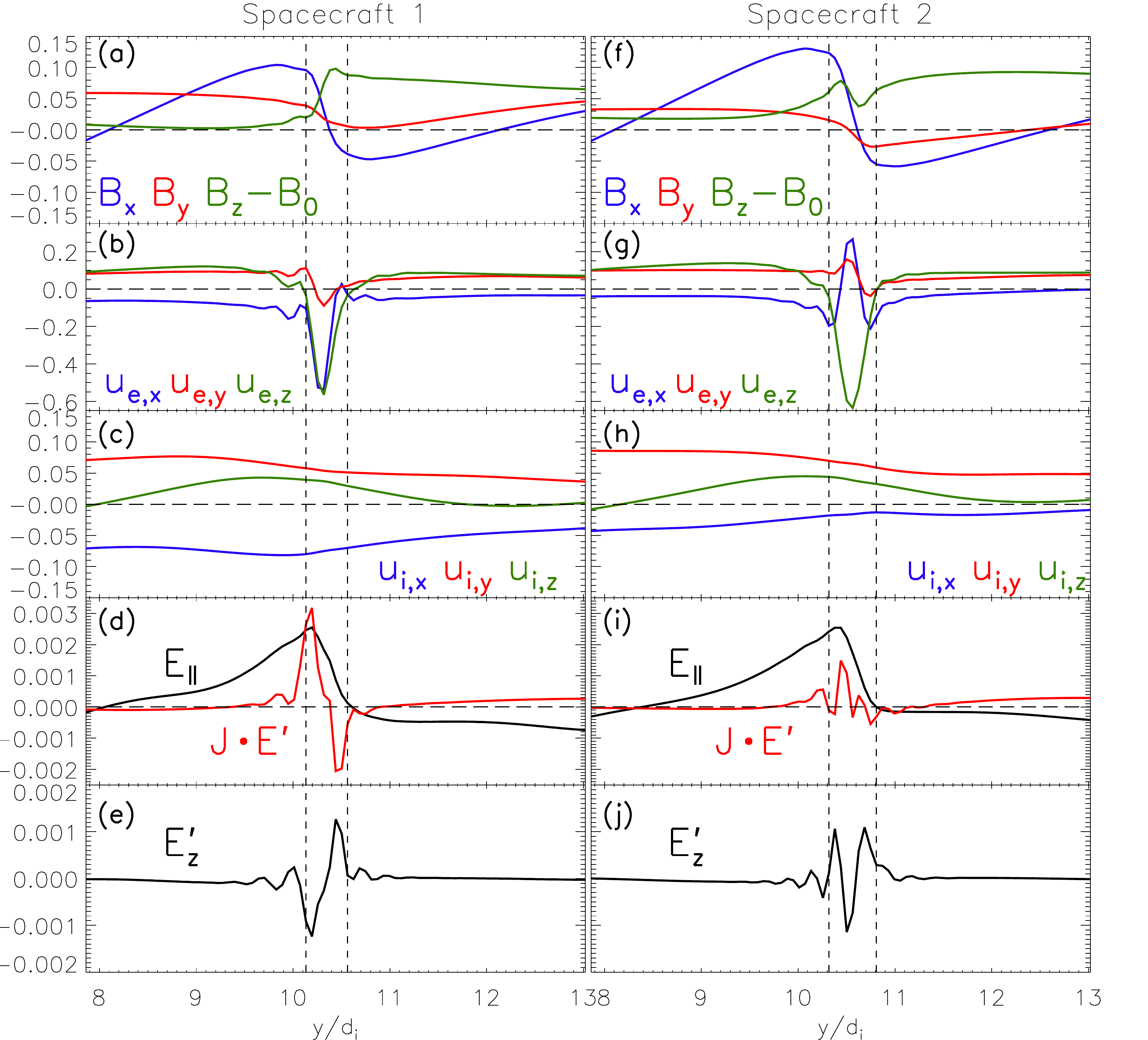}
\caption{From sim.A. Data taken by two virtual spacecraft passing through CS1 along the paths traced by the vertical dashed lines in figure \ref{fig2} (`Spacecraft 1' at $x\simeq57.4d_{\rm i}$, left column; and `Spacecraft 2' at $x\simeq58.4d_{\rm i}$, right column). The vertical dashed lines represent the local boundaries of the CS given by the condition $|J_z|>J_z^{\rm rms}$.
}\label{fig3} 
\end{figure*}

In figure \ref{fig3} we show data taken from two virtual spacecraft passing through CS1 along the paths traced by the vertical dashed lines in figure \ref{fig2}. These trajectories are chosen to be similar to those taken by {\it MMS} in \citet{PhanNAT2018}. In panels (a) and (f), we show the reversal of $B_x$ over a few $d_{\rm e}$, highlighted by the vertical dashed lines corresponding to the CS1 boundaries (given by the condition $|J_z|>J_z^{\rm rms}$). 
Within these boundaries, there are clear signatures of oppositely oriented super-Alfv\'enic electron jets, identified as the exhausts (panels (b) and (g)),  without any noticeable corresponding ion outflows (panels (c) and (h)). This feature is in qualitative agreement with {\it MMS} measurements~\citep{PhanNAT2018}. Note that ion jets do not appear even if we move the spacecraft trajectories further away from the X-point location, as would be the case in a `standard' reconnection event. The field-parallel electric field $E_\parallel$, $E_z'$, and the dissipation rate $\bb{J}\bcdot\bb{E}'$ \citep{zenitani2011} all depart significantly from zero only across the CS (panels (d), (e), (i), and (j)). Although $\bb{J}\bcdot\bb{E}'$ oscillates, its integral across CS1 is positive, indicating the possibility of a region of net magnetic-to-particle energy conversion (our equations only allow conversion into electron bulk, rather than thermal, energy). All of these features seen in CS1 are observed also in the other CSs.

\begin{figure}
\centering
\includegraphics[width=0.5\columnwidth]{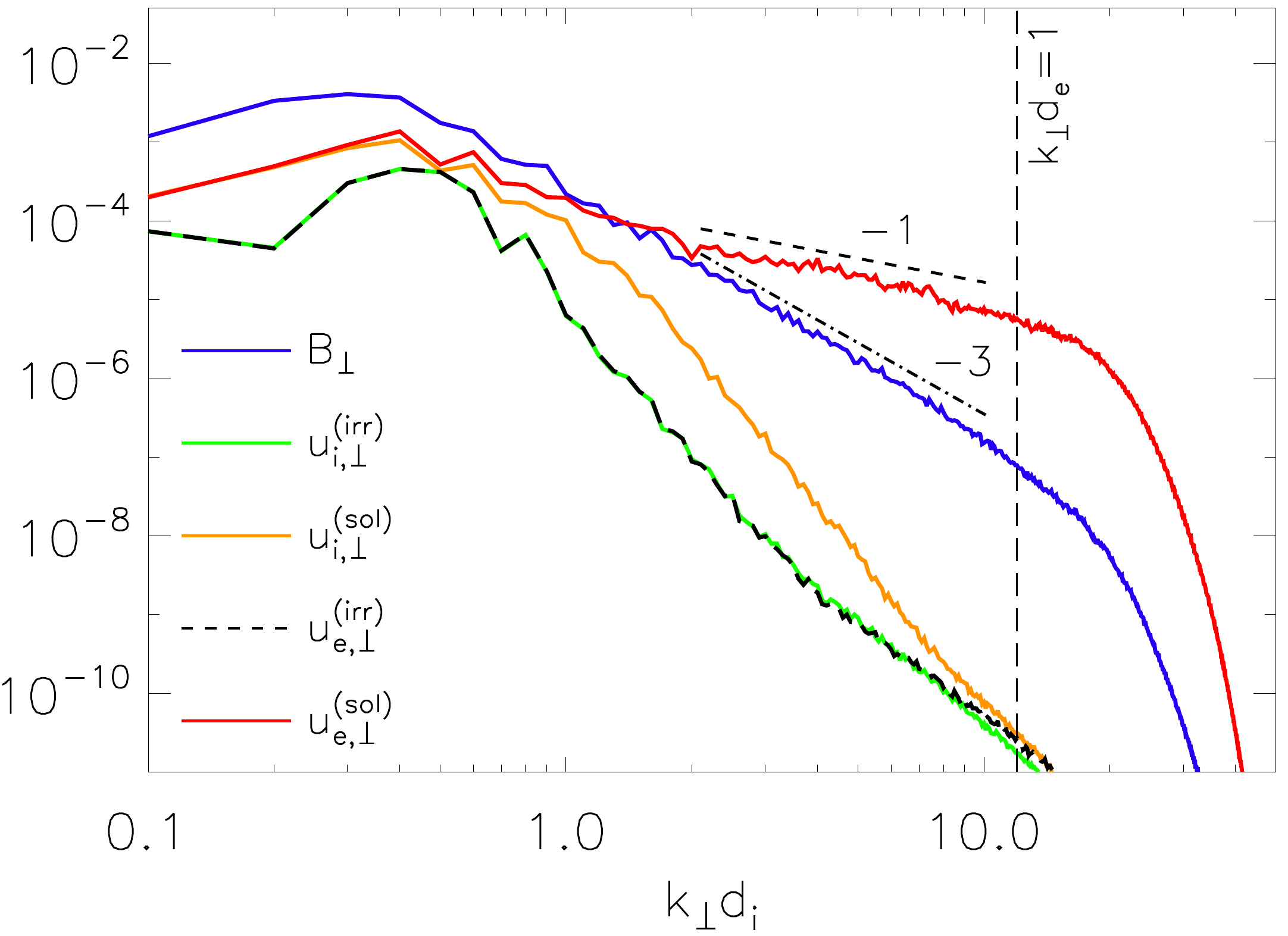}
\caption{From sim.A. Spectra at $t=42.5\Omega_{\rm i}^{-1}$ of the in-plane magnetic field (blue solid line), the solenoidal contributions to the in-plane ion and electron velocities $\bb{u}^{({\rm sol})}_{({\rm i,e})}$ (orange and red solid lines, respectively), and the irrotational contributions of the same velocities $\bb{u}^{({\rm irr})}_{({\rm i,e})}$ (green and black-dashed curves, respectively). Reference slopes of $-1$ and $-3$ slopes are provided.}\label{fig4} 
\end{figure}

That the dynamics of these CSs is EMHD-like is supported by figure \ref{fig4}, which shows spectra of the solenoidal $\bb{u}^{({\rm sol})}$ and irrotational $\bb{u}^{({\rm irr})}$ contribution to the in-plane ion and electron velocities and of the in-plane magnetic field during the onset of the first e-rec events ({\it viz.}, $t=42.5\Omega_{\rm i}^{-1}$). These spectra demonstrate that the ion flow is almost incompressible in the range $d_{\rm i}^{-1}<k_\perp<d_{\rm e}^{-1}$, while the electron flow remains nearly incompressible across an even larger range. We have also verified that the solenoidal contribution to $\bb{u}_{{\rm e},\perp}$ around the CSs largely dominates over its irrotational counterpart, and that the `EMHD terms' in our generalized Ohm's law dominate the dynamics ({\it viz.}, 
$|\grad\bcdot(n\bb{u}_{\rm i} u_{{\rm i}z})|\ll|\grad\bcdot(n\bb{u}_{\rm e} u_{{\rm e},z})|$ and $|n u_{{\rm e},z}(\grad\bcdot\bb{u}_{\rm e})|\ll|\bb{u}_{\rm e}\bcdot\grad(n u_{{\rm e},z})|$; see, e.g., \citealt{BulanovPOF1992}). 

Crucially, the properties of the fluctuations shown in figure \ref{fig4} at $t \sim 42.5\Omega_{\rm i}^{-1}$ do not change character once the fully developed turbulent stage is achieved at $t_{\rm p} \sim 110\Omega_{\rm i}^{-1}$ ($t_{\rm p}$ corresponds to the time when the rms out-of-plane current $J_z^{\rm(rms)}$ reaches its peak; see, e.g., \citealt{ServidioJPP2015}). After nearly 3 eddy turnover times, corresponding to the fully turbulent regime and by which time the ions could have begun participating in the reconnection dynamics, no sign of ion outflows in sim.A is observed. Figure \ref{fig5} (left panels) shows a comparison between the  out-of-plane and in-plane electron and ion flow velocities at $t=114\Omega_{\rm i}^{-1}$ in sim.A; the iso-contours of $-u_{{\rm e},z}\approx{J_{z}/n}$ (top left), $u_{{\rm i},z}$ (top right), $u_{{\rm e},y}$ (bottom left), and $u_{{\rm i},y}$ (bottom right) are shown. All electron quantities exhibit many thin structures at the electron scale, which are well correlated with the CSs traced by $J_z$ (not shown here, as it is nearly identical to $-u_{{\rm e},z}$). On the other hand, all of the ion quantities exhibit much smoother variations and, in particular, are largely uncorrelated with the corresponding electron flows or CSs. As in figure \ref{fig2}, electron jets are also visible (e.g., in $u_{{\rm e},y}$ at $(x,y)\approx(10,20)d_{\rm i}$; note that here the jets are along $y$), while no corresponding ion outflows are apparent. Despite these e-rec processes, the fluctuation spectra in the ion-kinetic range of the fully developed turbulence are not significantly affected (when compared to sim.B). The magnetic and electric energy spectra continue to exhibit sub-ion-scale power laws close to $-2.8$ and $-1$, respectively, in the range $d_{\rm i}^{-1} \lesssim k_\perp \lesssim d_{\rm e}^{-1}$ (not shown here), similar to those found in previous gyro-kinetic, hybrid-kinetic, and fully kinetic simulations \citep[e.g.,][]{HowesPRL2011,TenBargeAPJ2012,ToldPRL2015,FranciAPJ2015,FranciAPJ2016,FranciAPJ2018,CerriAPJL2016,CerriAPJL2017,CerriAPJL2018,GroseljAPJ2017,GroseljPRL2018,Arzamasskiy19,CerriFSPAS2019} as well as satellite measurements  \citep[e.g.,][]{AlexandrovaAPJ2008, SahraouiPRL2009, ChenPRL2010,LacombeAPJL2017}.\\

\subsection{Sim.B: Standard reconnection in kinetic turbulence}\label{sec:simBC}

None of the e-rec characteristics highlighted in \S\,\ref{sec:simA} are observed in sim.B, in which the magnetic perturbations are injected farther away from $d_{\rm i}$ (at least up to the time when the turbulence is fully developed, $t \approx 200\Omega_{\rm i}^{-1}$). In particular, as soon as CSs form in sim.B (after $1.5$ eddy-turnover times associated with the shortest initial wavelength, ${\sim}60\Omega_{\rm i}^{-1}$ here), we observe the development of `standard' reconnection
with electron structures embedded in ion macro-layers.
This is shown for sim.B in the top-right panel of figure \ref{fig1}, which provides a full-box view of the out-of-plane current density $J_z$ at $t = 95 \Omega_{\rm i}^{-1}$. In particular, we see the formation of a CS located at $x=18$, $y=50$ (highlighted by the black square). As in sim.A, the CS width collapses down to the $d_{\rm e}$ scale at the X-point. However, now its length along the in-plane magnetic field is much larger (more than $10 d_{\rm i}$) than those of the CSs observed in sim.A (left panels). As a consequence, at a sufficiently large distance from the X-point, ions couple to the magnetic-field/electron dynamics and ion outflows are generated. This is shown in the bottom-right panel (data versus $y$ taken along the dashed line): by way of comparison with its sim.A counterpart (bottom-left panel), the ions are seen to be better coupled to the magnetic field away from the CS. This facilitates the ions' participation to the reconnection process; indeed, a zoom-in of this CS (see figure \ref{fig6}) shows the X-point structure with both electron and ion outflows present. Even if smoother, the ion outflows point in opposite directions along the reconnecting in-plane magnetic field and are localized inside the separatrices.

To provide further evidence that the exhausts are composed of two coupled ion-electron jets, in figure \ref{fig7} we show six successive plots of the electron and ion outflows measured to the right of the X-point of the CS shown in figure \ref{fig6}. This sequence shows a central, well-collimated, strong electron jet (continuous lines) that becomes less collimated farther away from the X-point. We also observe a smoother, less intense but clear ion jet superposed on the electron one (dashed lines).

\begin{figure*}
\includegraphics[width=0.5\columnwidth]{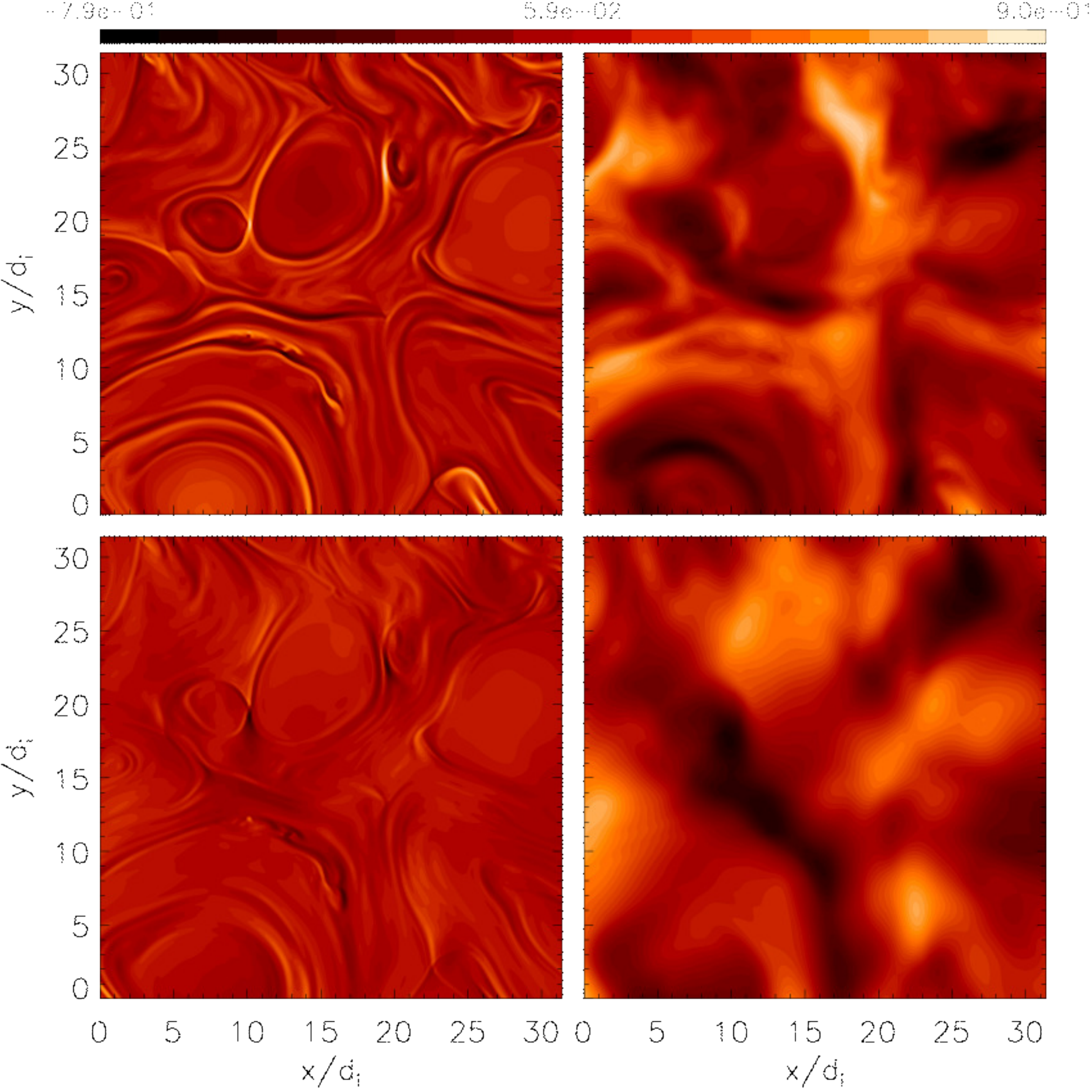}%
\includegraphics[width=0.5\columnwidth]{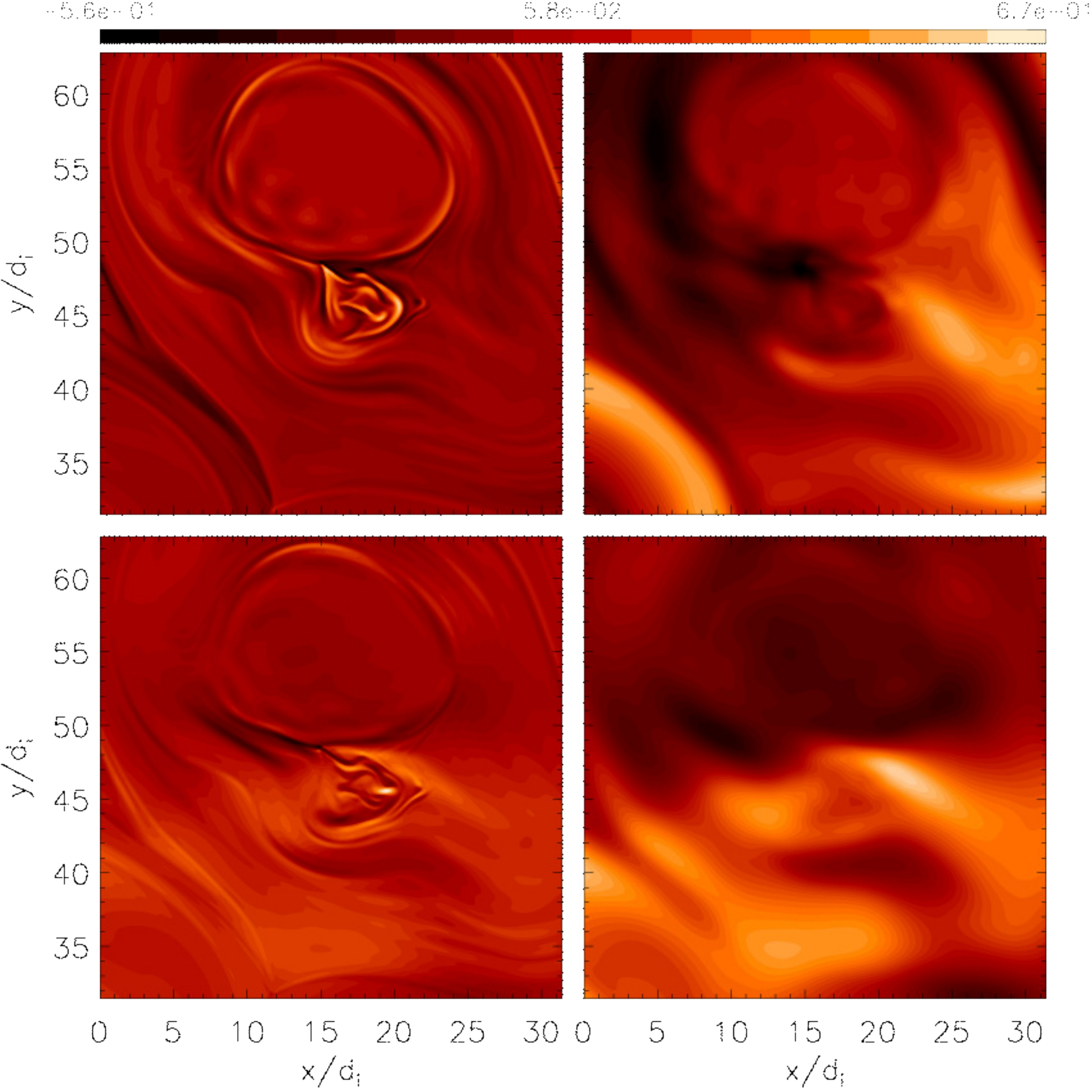}
\caption{Left: From sim.A. Shaded iso-contours of (minus) the out-of-plane electron flow, $-u_{{\rm e},z}\approx{J_z}/n$ (top left); the out-of-plane ion flow, $u_{{\rm i},z}$ (top right); and the $y$-component of the electron and ion flow, $u_{{\rm e},y}$ and $u_{{\rm i},y}$ (bottom left and bottom right, respectively) in a zoom-in region one-fourth the size of the simulation domain at $t=114\Omega_{\rm i}^{-1}$.
Right: From sim.B. Shaded iso-contours of (minus) the out-of-plane electron flow, $-u_{{\rm e},z}\approx{J_z}/n$ (top left); the out-of-plane ion flow, $u_{{\rm i},z}$ (top right); and the $x$-component of the electron and ion flow, $u_{{\rm e},x}$ and $u_{{\rm i},x}$ (bottom left and bottom right, respectively) in a zoom-in region one-fourth the size of the simulation domain at $t\approx203\Omega_{\rm i}^{-1}$.}\label{fig5}
\end{figure*}
\begin{figure*}
\includegraphics[width=0.5\columnwidth]{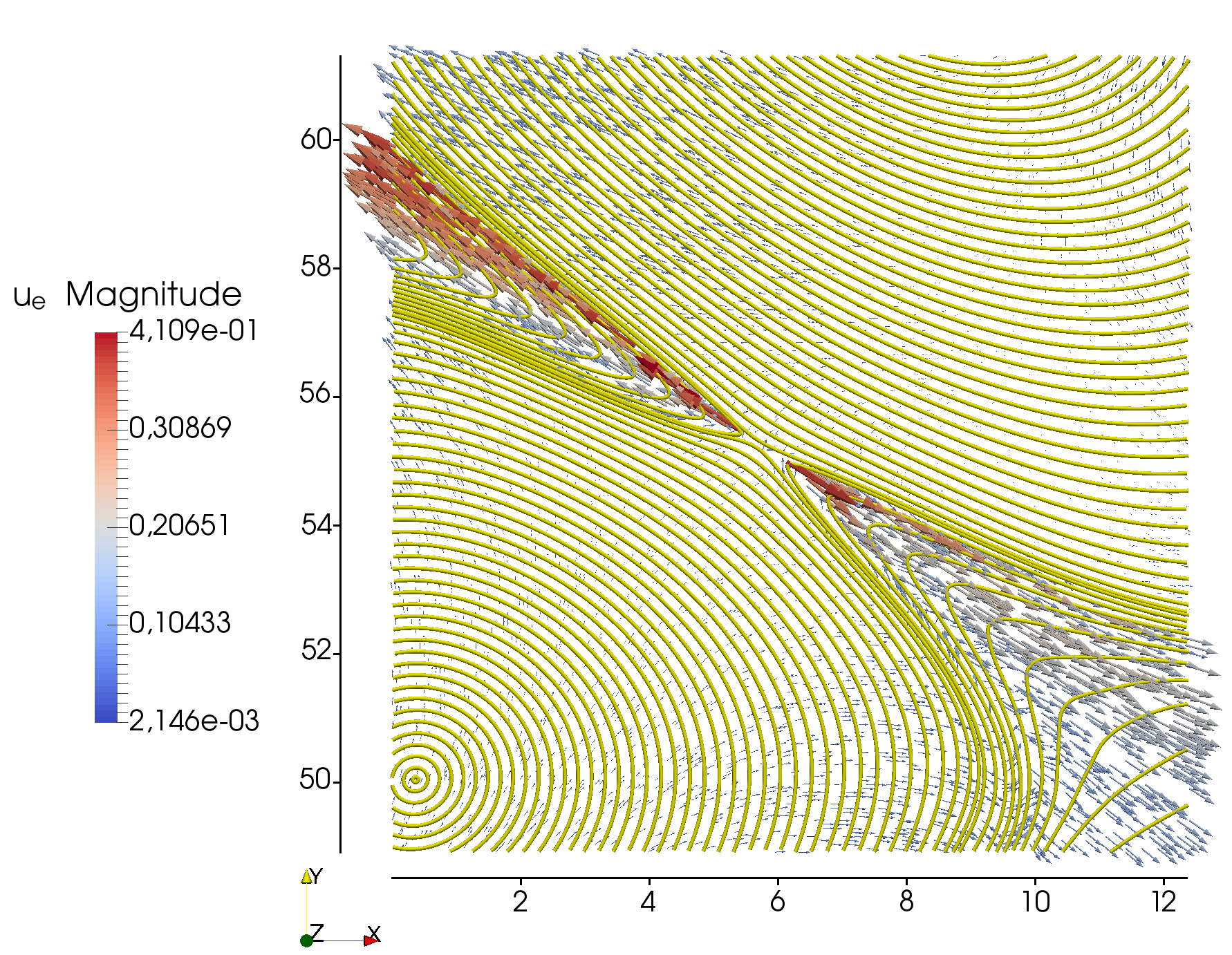}%
\includegraphics[width=0.5\columnwidth]{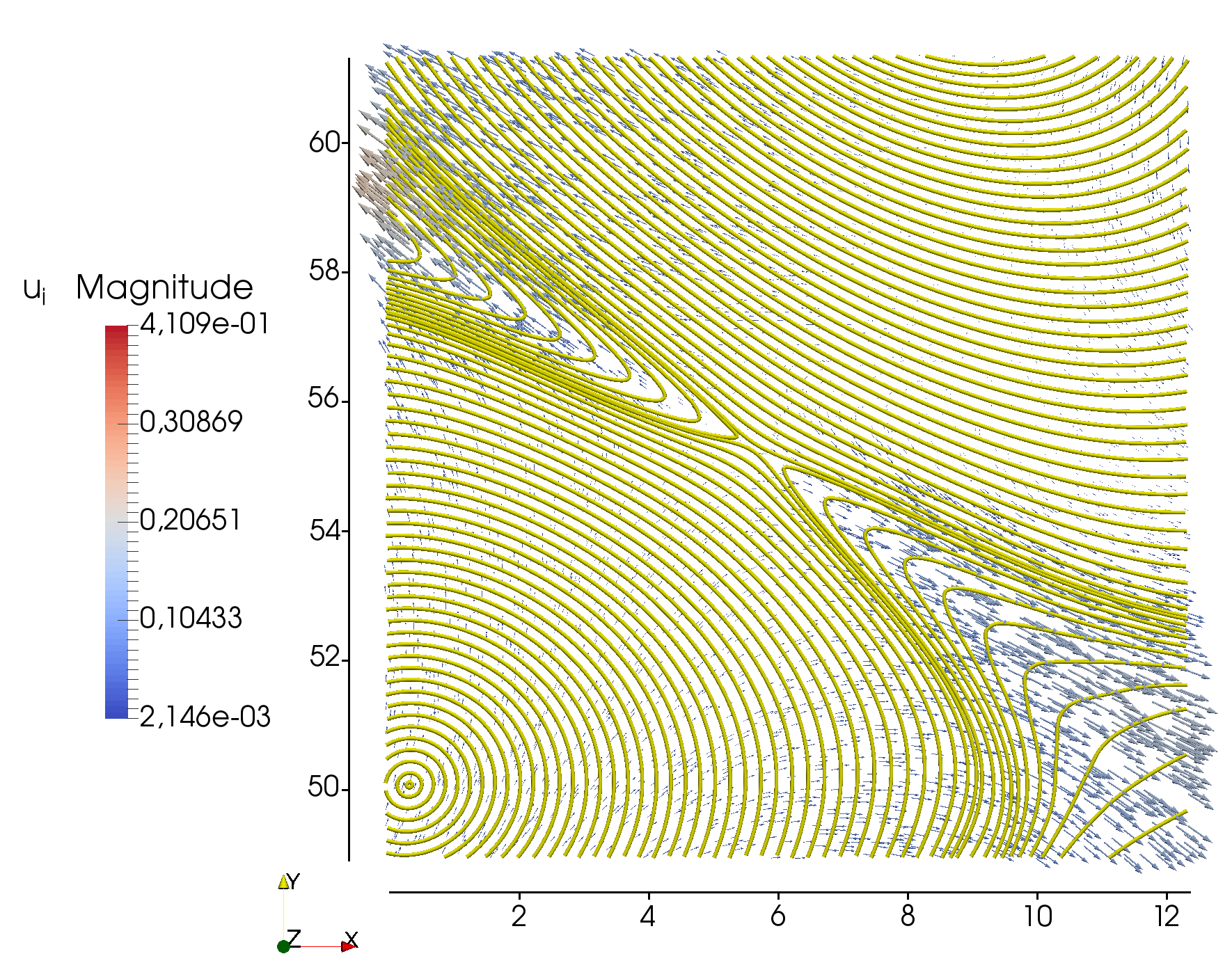}
\caption{From sim.B. In-plane velocity vector field for electrons (left) and ions (right) with superimposed the flux function $\psi$ iso-contours around the X-point at $x \approx 6d_{\rm i}$, $y \approx 56d_{\rm i}$, at $t=105\Omega^{-1}_{\rm i}$.}\label{fig6} 
\end{figure*}
\begin{figure*}
\centering
\includegraphics[width=\columnwidth]{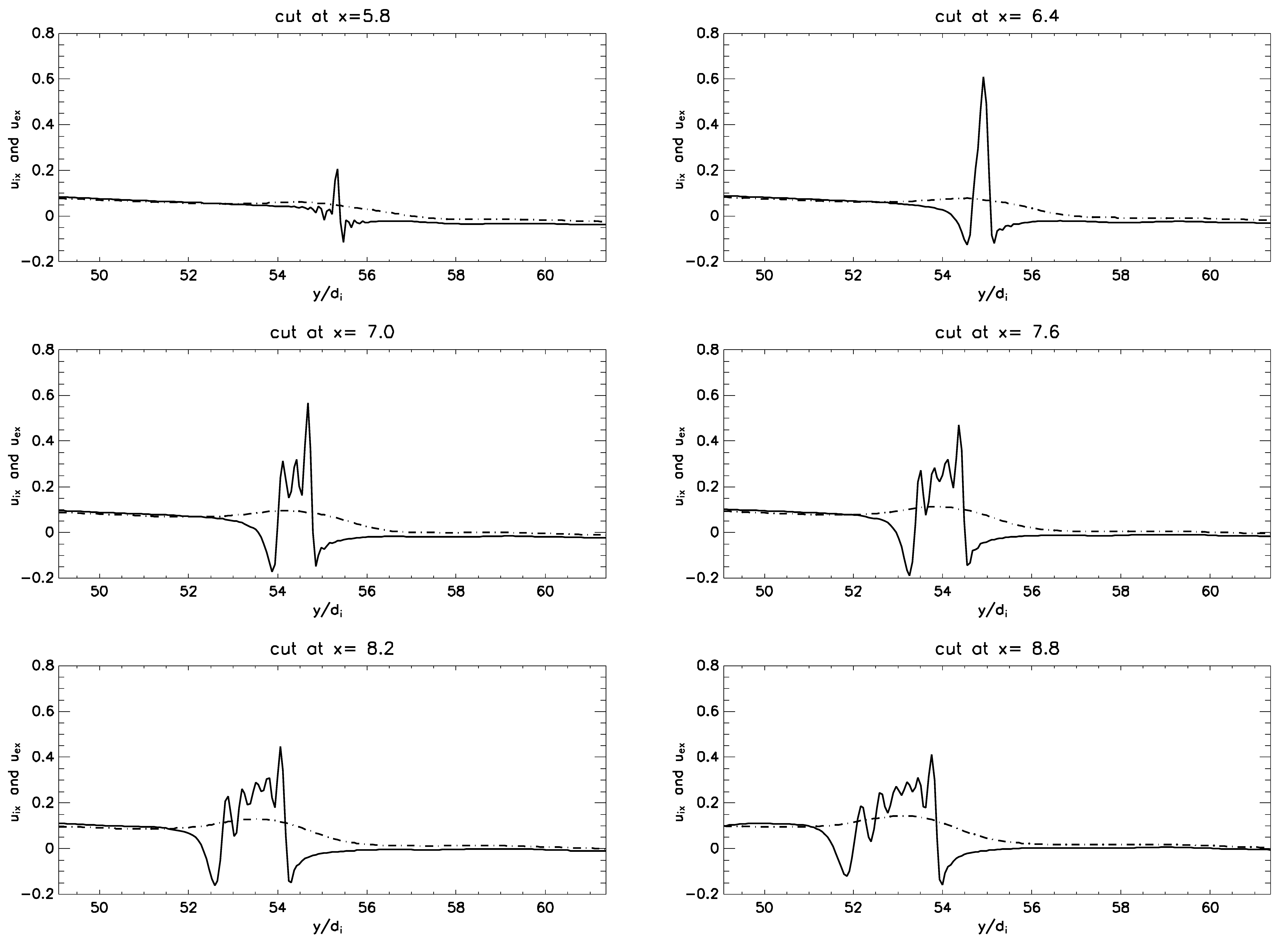}
\caption{From sim.B. Cuts versus $y$ of the rightward outflow from the X-point shown in figure \ref{fig6}, at $t = 105 \Omega_{\rm i}^{-1}$. The $x$-component of the electron (solid) and ion (dash-dotted) velocities are plotted versus $y$ at different values of $x$ starting close to the X-point and going as far away as $3 d_{\rm i}$.}\label{fig7} 
\end{figure*}
Such an electron-ion structure is never observed in sim.A where the ions decouple from the magnetic-field and the electron dynamics over nearly the entire simulation domain (see figure \ref{fig1}, bottom left frame, and  figure \ref{fig5}, left frame). This difference is observed not only when the first active CSs form, but also at later times when the turbulence is fully developed. We also note that, although the ion-electron coupling remains valid in the fully developed turbulence stage of sim.B, some electron-scale structures
emerge, as shown in figure \ref{fig5}, right frame, where finer structures develop inside an ion scale reconnecting structure. 
In summary, in sim.B the different CSs exhibit `standard' {or even} multi-scale reconnection with both {well-developed} ion and electron outflows. On the contrary, in sim.A only e-rec develops and `standard' reconnection is never observed, neither at the time when the first CSs activate nor when the turbulence is fully developed.\\

\subsection{Statistical analysis of current sheets' properties}\label{sec:manu}

Quantitative differences between sim.A and sim.B {can be further assessed} by examining the statistical distribution of the CSs' characteristic widths and lengths. For this purpose we define a CS as a region where the magnitude of the current density is bigger than a {given} threshold $J_{\rm thr}$. {Following} 
\citet{ZhdankinAPJ2013} we define such a threshold as $J_{\rm thr} = \sqrt{<J_z^2>\, +\, 3\,\sigma}$, where $\sigma = \sqrt{<J_z^4> - (<J_z^2>)^2  }$. The widths and lengths of the CSs are then evaluated using two different techniques. 

The first technique employs an automated procedure that calculates the width and length of each CS based on the shape of the local $|J_z|$. First, we define the (local) current peak as the maximum value of $|J_z|$. Then, given the eigenvector associated with the largest eigenvalue of the Hessian matrix of $|J_z|$ at the peak, we look at the interpolated profile along this direction and define the CS's width as the full width at half maximum of $|J_z|$. Performing a similar procedure to compute the length, namely, interpolating $|J_z|$ along the direction of the eigenvector associated with the smallest eigenvalue, could be misleading since the CSs are rarely `straight'. For this reason we infer the length as the maximal distance between two points belonging the same CS (i.e., within the thresholded contour), following {\citet{ZhdankinAPJ2013}}.

\begin{figure*}
\centering
\includegraphics[width=0.7\columnwidth]{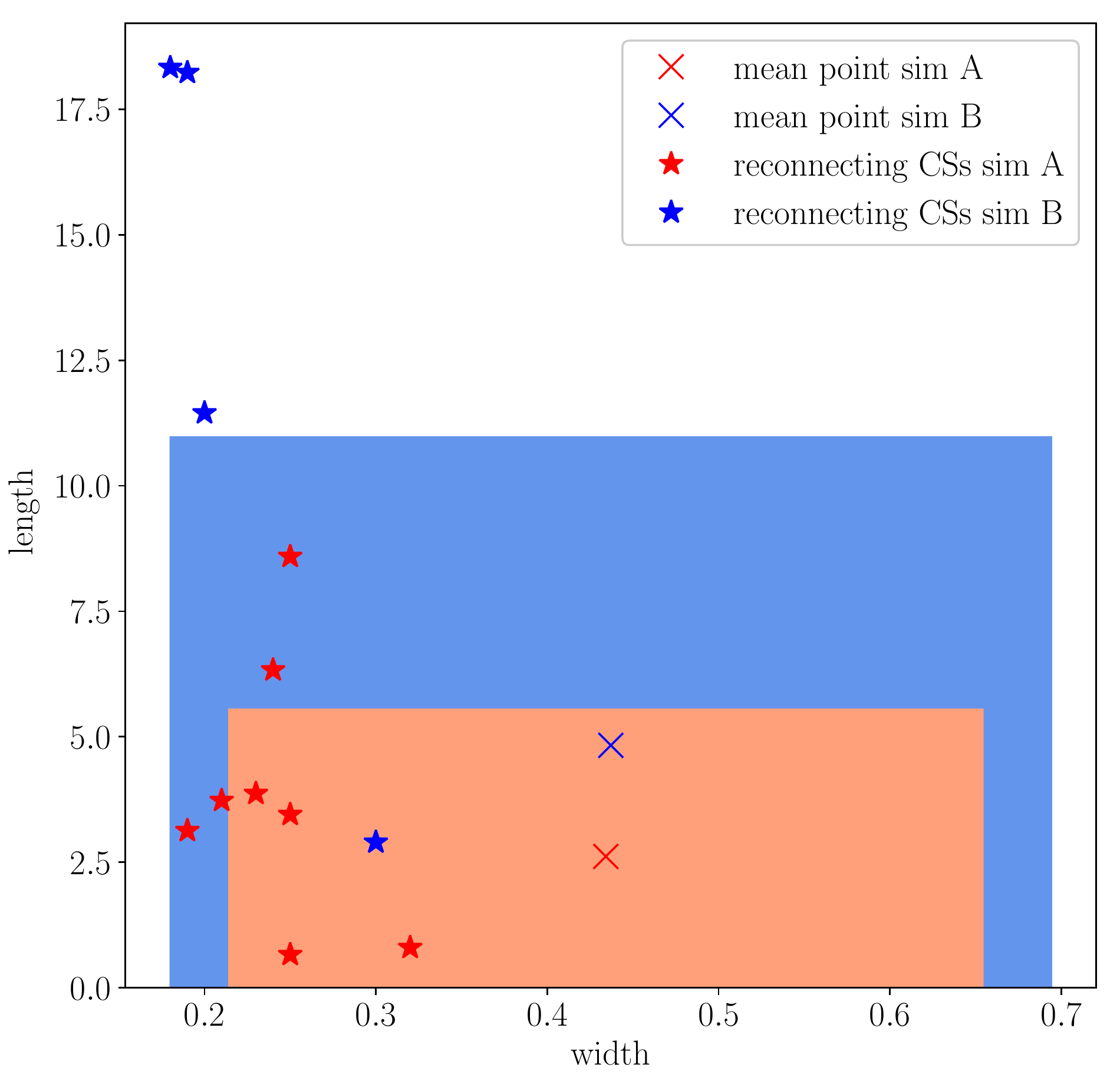}
\caption{A visual representation of the statistical distribution of width and length of all the CSs present in sim.A (red) and sim.B (blue): the cross indicates the average values and the shaded area the parameter region within one standard deviation. The widths and lengths of selected CSs that exhibit an X point and are clearly reconnecting are indicated by the stars.}\label{fig8} 
\end{figure*}

We then evaluate the average CS width and length and their standard deviations by taking into account all of the current peaks present in a simulation between the time when the first CSs form ($t=30\Omega_{\rm i}^{-1}$ in sim.A and $t=75\Omega_{\rm i}^{-1}$ in sim.B) and the fully developed turbulent regime ($t=128\Omega_{\rm i}^{-1}$ in sim.A and $t=203\Omega_{\rm i}^{-1}$ in sim.B); in total, 7182 (460) current peaks were identified in sim.A (sim.B). These quantities are shown in figure \ref{fig8} for both sim.A (red markers) and sim.B (blue markers):  the crosses indicate the average width/length of these runs, while the red (blue) shaded area indicates the parameter region within one standard deviation of these values for sim.A (sim.B).
Note that the average width (a few $d_{\rm e}$) and its standard deviation are very similar amongst sim.A and sim.B, suggesting that these characteristics are independent of the details of the energy injection. This is not particularly surprising, since the minimum width of a CS width is limited by the decoupling of the magnetic field from the electron dynamics. 
By contrast, the length distributions are noticeably different in sim.A and sim.B. The average CS length in sim.A is ${\simeq}2.6d_{\rm i}$ and the standard deviation is ${\simeq}2.9d_{\rm i}$, while in sim.B these values correspond to ${\simeq}4.8d_{\rm i}$ and ${\simeq}6.2d_{\rm i}$, respectively. Therefore CSs in sim.B are not only longer on average, but they also show a more pronounced variability in length.
The picture emerging from sim.B therefore conforms to the `standard' idea of plasma turbulence developing a hierarchy of dissipative coherent structures extending down to electron scales'~\citep{KarimabadiPOP2013}, whereas sim.A provides the first numerical evidence of the possibility to `shortcut' this hierarchy and develops only smaller-scale structures where e-rec occurs.

This first analysis procedure has the advantage of being automated, leading to a large statistical ensemble of CSs. That being said, this procedure does not distinguish between reconnecting CSs and non-reconnecting ones. For example, CSs across which the in-plane magnetic field (the only one that can reconnect in 2D) does not change sign but whose current densities are sufficiently large are counted by the automated procedure, despite their inability to reconnect. There are also CSs across which the in-plane magnetic field changes sign, but no clear signs of reconnection (e.g., electron and/or ion outflows emerging from an X point) are present. Our second technique is therefore to compute the widths and lengths of only those CSs that exhibit clear evidence of reconnection. At a fixed time, we visually inspect each CS (identified by our first technique) to assess whether the in-plane magnetic field changes sign and whether electron outflows are present. We then examine those CSs' time evolution to find evidence for the erosion of $\psi$ at the X point, indicating an ongoing topological modification of the magnetic field. The measured widths and lengths of these reconnecting CSs are marked by stars in figure \ref{fig8}. We have verified, using the proxies $J_z$ and $E_z^{'}$, that the estimated widths of these CSs  represented by a star are correct.

Note that the widths of these
reconnection sites are always smaller than the average CSs' width, in both sim.A and sim.B, corroborating the fact that active CSs collapse at the X point. By contrast, the lengths of reconnecting CSs exhibit quite different behavior amongst these two simulations: in sim.A the majority of reconnecting CSs have a length in between the boundaries defined by the standard deviation, with only {two cases (out of $8$)}  {of reconnecting CSs with a} {slightly larger}
length (although smaller than $9d_{\rm i}$). On the other hand, all but one reconnecting CSs in sim.B are out of the boundaries defined by the standard deviation and have a length larger than $10 d_{\rm i}$. The only one with a smaller length, at ${\simeq}2.9 d_{\rm i}$, is the CS found at $t=203\Omega_{\rm i}^{-1}$ where fine electron structures are visible (see Figure \ref{fig5}).

Even if the number of these reconnection sites is statistically smaller than the total number of CSs, we can identify a trend regarding CS width and length within the two simulations. In general, the ensemble of all the CSs in a simulation is not particularly representative of the reconnecting ones. In fact, all the reconnecting CSs have a width that is smaller than the average value of the corresponding distribution of CSs. This width is clustered around $\sim(0.2$--$0.25)d_\mathrm{i}\approx3d_\mathrm{e}$ in both simulations, with a relatively small dispersion around that value (the width being always less than $0.35d_\mathrm{i}$).
More interestingly, the lengths of the reconnecting CSs exhibit a completely different trend amongst the two simulations. In sim.A, the length of these reconnecting structures (which exhibit e-rec) is clustered around $\sim 3.5d_\mathrm{i} \approx 42 d_\mathrm{e}$, and nearly all of them have a characteristic length which is within one standard deviation from the average value of the distribution of all the CSs in sim.A. In sim.B, on the other hand, nearly all of the reconnecting CSs (which exhibit ion outflows) have a length that is beyond one standard deviation from the average value of the distribution of all the CSs, i.e., larger than $10d_\mathrm{i}=120d_\mathrm{e}$.

\section{Conclusions and Discussion}

Using numerical simulations of kinetic plasma turbulence in a hybrid Vlasov--Maxwell model that includes fully non-linear electron-inertia effects (except for those related to the electron pressure tensor terms), we have demonstrated for the first time that short CSs in which electron-only reconnection takes place can naturally develop within kinetic plasma turbulence at $\beta\sim1$ when fluctuations are injected on a range of wavenumbers near ion-kinetic scales (in our case, $k_\perp d_{\rm i} = k_\perp \rho_{\rm i} \le 0.6$). In this situation, all CSs showing clear evidence of ongoing magnetic reconnection exhibit electron exhausts that are unaccompanied by ion outflows, with the ions being decoupled from the magnetic-field and electron dynamics almost everywhere. The properties of these reconnecting CSs, including their statistical lengths and widths are in qualitative agreement with those ``electron-scale'' CSs recently measured in the turbulent magnetosheath by {\it MMS} \citep{PhanNAT2018}, in which reconnection-driven electron outflows were unaccompanied by ion outflows. 
It is worth noticing that in this case where electron reconnection occurs associated with smaller CSs than for standard reconnection (sim.A), we do not observe the classical 2D island coalescence process \citep{malara92}, at least during the full duration of the simulation up to $t \simeq 140$.

In our simplified model in which electrons are modeled as a fluid and decaying turbulence is initialized via compressive magnetic perturbations, this transition from `standard' to `electron-only' dynamics is observed by varying the maximum wave number of the initial fluctuations. In particular, if fluctuations are injected at wavelengths such that $k_\perp d_{\rm i} \le 0.3$, a hierarchy of dissipative CSs emerge with a variety of lengths, whose evolution follows the `standard' reconnection dynamics in which both ion and electron exhausts are seen.

We have further shown that the wavelength range of energy injection has a measurable impact on the statistical properties of all the CSs: their average length and corresponding standard deviation are indeed different in sim.A and sim.B. In particular, sim.A shows a distribution of CSs' lengths that is centered around ${\simeq}2.6 d_{\rm i}$ with a standard deviation of their (asymmetric) distribution of ${\simeq}2.9 d_{\rm i}$, whereas the corresponding distribution in sim.B exhibits a larger average value (${\simeq}4.8 d_{\rm i}$) and a much wider distribution, with a standard deviation of ${\simeq}6.2 d_{\rm i}$. This difference is even more apparent if we limit the analysis to only those CSs that can be clearly identified as reconnecting ones:
in sim.A, these CSs are mostly clustered around a length of ${\simeq}3.5d_{\rm i}$ and are never longer than $\approx 8$--$9 d_{\rm i}$. Moreover, their distribution is relatively thin and similar to the one obtained using all the CSs. By contrast, in sim.B we find both very long reconnecting CSs (up to ${\sim}20 d_{\rm i}$) and short ones (a few $d_{\rm i}$). These statistical properties directly relate to the physics of reconnection that is observed: short CSs in sim.A only produce e-rec, while long and short CSs in sim.B allow for the presence of ion outflows and finer electron structures. 
Note that, differently to reconnection in a 1D initial Harris sheet \citep{fujimoto_pop_2006,Daughton_pop_2006} where the length of the resulting CS is not limited in size and grows in time following the initial configuration, in the turbulent case CSs' characteristic length is limited. Indeed CSs develop in between magnetic flux ropes whose size sets the maximal length CSs can reach. Therefore, the energy injection scale plays an important role by setting the size of flux ropes, and so on the CSs' maximal length and on the possible kind of reconnection developing. Recent fully kinetic simulations of magnetic reconnection developing in an isolated current sheet~\citep{SharmaPyakurel2019} are in agreement with this picture.

Despite our initial conditions being oversimplified and not directly tied to any specific physical injection mechanism, we are encouraged by the close resemblance of sim.A's results to {\it MMS} data. That this resemblance holds not only during the onset of the first reconnection events in sim.B, but also during the fully developed turbulent state, is particularly convincing. We are then led to conjecture that energy injection occurring near ion-kinetic scales, perhaps due to velocity-space instabilities and/or shocks, can qualitatively alter the evolution of CSs and the dynamics of magnetic reconnection in plasma turbulence, potentially explaining the {\it MMS} results. 
In particular, if some energy-injection mechanism with properties similar to those employed in our simulations occurs past the bow shock, our results suggest that e-rec events could be detected relatively close to the bow shock and also further downstream in the magnetosheath. A more recent observational study using an extended {\it MMS} data set from the turbulent magnetosheath~\citep{StawarzAPJL2019} seems to support this scenario. As the {\it MMS} mission goes on, the amount of collected data will enable a more detailed statistical analysis of these e-rec events. This kind of analysis should include also data collected outside of the magnetosheath, in the pristine solar wind, where different statistics for the reconnecting current sheets may be expected. However, due to the different level of fluctuations in the two regions, we stress that a high sensitivity of the spacecraft's instruments may be required in order to detect (the presumably smaller amount of) e-rec events in the pristine solar wind. Next-generation space missions would be most informative.

\section*{Conflict of Interest Statement}
The authors declare that the research was conducted in the absence of any commercial or financial relationships that could be construed as a potential conflict of interest.

\section*{Author Contributions}

All authors have contributed to the data analysis, text editing and to the discussions presented in the paper.

\section*{Funding}
This project (FC) has received funding from the European Union's Horizon 2020 research and innovation programme under grant agreement No 776262 (AIDA). 
SSC and MWK were supported by the National Aeronautics and Space Administration under Grant No.~NNX16AK09G issued through the Heliophysics Supporting Research Program.

\section*{Acknowledgments}
We acknowledge Referee 3 for her/his comments and discussion helping us to improve the content of the paper. We acknowledge PRACE for awarding us access to the supercomputer Marconi, CINECA, Italy, where the calculations were performed under the grant n.~2017174107.  FC and SSC thank Dr.~C.~Cavazzoni (CINECA, Italy) for his essential contribution to code parallelization and performance and Dr.~M.~Guarrasi (CINECA, Italy) for his contribution on code implementation on Marconi. This work was also partly supported by CINES/IDRIs under the allocation 2018-A0040407042 made by GENCI.

\section*{Data Availability Statement}
The data used in this study are available from the authors upon reasonable request.


\end{document}